\documentclass[amsmath,amssymb,preprintnumbers,nofootinbib,prd,a4paper,twocolumn]{revtex4-1}
\pdfoutput=1
\usepackage{amsthm}
\usepackage{amsmath}
\usepackage{graphicx}
\usepackage{bbold}
\usepackage{color}
\usepackage{subfigure}
\usepackage{multirow}
\usepackage{graphicx,bm}

\usepackage{dcolumn}% Align table columns on decimal point
\usepackage{bm}% bold math
\usepackage{slashed}
\usepackage{epsfig}
\usepackage{hyperref}
\usepackage{verbatim}

\usepackage[utf8]{inputenc}
\usepackage[T1]{fontenc}
\usepackage[normalem]{ulem}

\newcommand{\beq}{\begin{eqnarray}}
\newcommand{\eeq}{\end{eqnarray}}

\newcommand{\bmp}{\noindent\begin{minipage}{16cm}}
\newcommand{\emp}{\end{minipage}\vskip 7mm} % 7mm untightened

\newcommand{\be}{\begin{eqnarray}}
\newcommand{\ee}{\end{eqnarray}}

\begin{document}
%%%%%%%%%%%%%%%%%%%%%%%%%%%%%%%%%%%%%%%%%%%%%%%%%%%%%%%%%%%%%%%%%%%%%%%%%%%
\title{XENON1T solar axion and the Higgs boson emerging from the dark}
%%%%%%%%%

\author{Chengfeng {\sc Cai}}
%\email{}
\affiliation{School of Physics, Sun Yat-Sen University, Guangzhou 510275, China}
\author{Giacomo {\sc Cacciapaglia}}
\email{g.cacciapaglia@ipnl.in2p3.fr}
\affiliation{Institut de Physique des 2 Infinis (IP2I),
CNRS/IN2P3, UMR5822, 69622 Villeurbanne, France}
\affiliation{Universit\' e de Lyon, Universit\' e Claude Bernard Lyon 1, 69001 Lyon, France}
\author{Martin {\sc Rosenlyst}}
\email{rosenlyst@cp3.sdu.dk}
\affiliation{CP$^3$-Origins, University of Southern Denmark, Campusvej 55, DK-5230 Odense M, Denmark}
\author{Hong-Hao {\sc Zhang}}
\email{zhh98@mail.sysu.edu.cn}
\affiliation{School of Physics, Sun Yat-Sen University, Guangzhou 510275, China}
\author{Mads T. {\sc Frandsen}}
\email{frandsen@cp3.sdu.dk}
\affiliation{CP$^3$-Origins, University of Southern Denmark, Campusvej 55, DK-5230 Odense M, Denmark}

%%%%%
%%%%%%%%%%%%%%%%%%%%%%%%%%%%%%%%%%%%%%%%%%%%%%%%%%%%%%%%%%%%%%%%%%%%%%%%%%%

\begin{abstract}
In a recent letter we proposed a new non-thermal mechanism of Dark Matter production based on vacuum misalignment, where both the Higgs boson and a very light pseudo-scalar $\eta$ emerge from the Dark sector. In this letter, we identify the parameter space in a composite scenario where the light pseudo-scalar can be produced in the sun and explain the XENON1T excess in electron recoil data. The model's Dark Matter candidate has a mass around $50$~TeV and out of range for Direct Detection. Testable predictions include Gravitational waves at frequencies in the Hz range from a cosmological phase transition, an exotic decay $Z \to \gamma + \mbox{inv.}$ with rates $4 \div 16 \cdot 10^{-12}$ testable at a future Tera-Z collider, and an enhancement by $17\div 40$\% of the branching ratio $K_L \to \pi^0 + \mbox{inv.}$, not enough to explain the KOTO anomaly. All these predictions may be confirmed by future experiments.

\end{abstract}

\maketitle

%%%%%%%%%%%%%%%%%%%%%%%%%%%%%
%\section{Introduction}

In a recent letter~\cite{Cai:2019cow} we have presented a novel mechanism that can produce the needed Dark Matter relic density
in a non-thermal way. This framework applies when the electroweak symmetry breaking is due to  vacuum misalignment and the 
Higgs boson emerges as a pseudo-Nambu-Goldstone boson (pNGB). A large class of models of New Physics are thus eligible, ranging
from models based on strong dynamics, like composite Higgs~\cite{Kaplan:1983fs,Dugan:1984hq} and Little Higgs~\cite{ArkaniHamed:2001nc,ArkaniHamed:2002qx} models, to weakly coupled models,
like holographic extra dimensions~\cite{Contino:2003ve,Hosotani:2005nz} and elementary Goldstone boson models~\cite{Alanne:2014kea}.
The kernel of the Dark Matter production mechanism lies in the thermal history of the cooling Universe:
at high temperature, the vacuum of the model consists of an essentially
Higgsless phase, where the electroweak symmetry is broken at a scale $f \gg v_{\rm SM} = 246$~GeV, and a global U(1)$_X$ allows
for the generation of an asymmetric density of pNGBs deriving from a large global symmetry,
spontaneously broken by the vacuum. At a temperature $T_\ast < f$, the vacuum starts rotating away from the Higgsless direction and 
towards an alignment where the Higgs boson can be identified with a light pNGB of the theory, while the U(1)$_X$ is spontaneously broken.
Part of the asymmetry is thus stored in stable $\mathbb{Z}_2$--odd pNGBs, the lightest of which plays the role of the Dark Matter
currently filling in the Universe. In the Higgsless vacuum, the would-be Higgs boson $h$ forms a U(1)$_X$ charged scalar, $\phi_X = (h + i \eta)/\sqrt{2}$,
with a neutral pseudo-scalar $\eta$. A generic prediction of the model is that $\eta$ remains very light in the zero-temperature vacuum.
Thus, both the Higgs boson and a very light pseudo-scalar emerge from the Dark sector of the theory.

It's tantalising  that on June 17, 2020, the XENON1T collaboration unveiled the presence of an excess in the electron recoil events~\cite{Aprile:2020tmw},
which may be compatible with the production of an axion-like particle (ALP) in the sun. The mass of this new pseudo-scalar should be
below $100$~eV to explain the data. In this letter we will ask the question: can the pseudo-scalar $\eta$, emerging from the Dark, explain this excess?

While the Dark Matter production mechanism of Ref.~\cite{Cai:2019cow} applies to a large variety of models, to study the XENON1T excess 
we will focus on the possibility that 
the dynamics is based on compositeness. The advantage of this approach is that the couplings of the $\eta$ to gauge bosons can be
predicted in terms of the Wess-Zumino-Witten (WZW) topological anomaly~\cite{Wess:1971yu,Witten:1983tw}, while couplings to fermions need to be negligible in order to
preserve the U(1)$_X$ symmetry. The minimal global symmetry breaking patterns that can accommodate the Dark Matter production
mechanism are SU(6)/Sp(6) and SU(4)$\times$SU(4)/SU(4).~\footnote{In both cases, the $\mathbb{Z}_2$ stable states may also be standard freeze-out thermal candidates~\cite{Frigerio:2012uc}, see~\cite{Ma:2017vzm,Cai:2018tet}.} In both cases, the WZW couplings of $\eta$ can be written as~\cite{Galloway:2010bp,Ma:2015gra}
\begin{equation} \label{eq:WZW}
\mathcal{L}_{\rm WZW} = \frac{C_{\rm WZW}}{\Lambda} \eta \left( g^2 W^a_{\mu \nu} \tilde{W}^{a,\mu \nu} - {g'}^2 B_{\mu \nu} \tilde{B}^{\mu \nu} \right)\,,
\end{equation}
where $W^a$ ($B$) indicate the gauge bosons of SU(2)$_L$ (U(1)$_Y$) with gauge coupling $g$ ($g'$), and $\tilde{G}^{\mu \nu} = \frac{1}{2} \epsilon^{\mu \nu \rho \sigma} G_{\rho \sigma}$. The coefficient can be expressed as follows:
\begin{equation} \label{eq:CWZW}
\frac{C_{\rm WZW}}{\Lambda} = \frac{d_\psi \cos \theta}{64 \sqrt{2} \pi^2 f}\,,
\end{equation}
where the composite Higgs decay constant $f$ is related to the electroweak scale $v_{\rm SM}$ via the misalignment angle $\theta$, as $f\sin \theta = v_{\rm SM}$.
Moreover, $d_\psi$ contains information about the underlying gauge dynamics leading to the formation of the composite pNGBs: namely, $d_\psi$ is the dimension of the representation of the fermions $\psi$, which form the pNGBs, under the confining hypercolour gauge symmetry $\mathcal{G}_{\rm HC}$. 
For the coset SU(6)/Sp(6), the minimal model is based on $\mathcal{G}_{\rm HC} = \mbox{SU(2)}_{\rm HC}$ with $\psi$ in the fundamental ($d_\psi = 2$)~\cite{Galloway:2010bp,Cacciapaglia:2014uja}, while top partial compositeness can be obtained for $\mathcal{G}_{\rm HC} = \mbox{Sp(4)}_{\rm HC}$ ($d_\psi = 4$)~\cite{Barnard:2013zea,Ferretti:2013kya}. For SU(4)$\times$SU(4)/SU(4), the minimal model is based on $\mathcal{G}_{\rm HC} = \mbox{SU(3)}_{\rm HC}$ with $\psi$ in the fundamental ($d_\psi = 3$)~\cite{Vecchi:2015fma,Ma:2015gra}, while top partial compositeness can be generated for $\mathcal{G}_{\rm HC} = \mbox{SU(4)}_{\rm HC}$ with $\psi$ in the fundamental ($d_\psi = 4$)~\cite{Ferretti:2013kya,Cacciapaglia:2018avr}. In the following, we will
prefer models with top partial compositeness, and fix $d_\psi = 4$ as the minimal choice. Once the underlying dynamics is fixed, the interactions of $\eta$ only depends on $f$ \footnote{Changing $d_\psi$ can be absorbed in a simple rescaling of $f$, as evident in Eq.~\eqref{eq:CWZW}.}, as $\cos \theta \sim 1$ due to precision bounds~\cite{Grojean:2013qca,Ghosh:2015wiz}. The other remaining free parameter is the mass. The system is, therefore, very constrained and 
the XENON1T excess offers a golden chance to test it against data.

%%%%%%%%%%%%%%%%%%%%%%%%%%%%%%%%%%%
\section{The model in the XENON1T arena}

At energies well below the electroweak scale, the $\eta$ couplings match those of a generic ALP~\cite{Bauer:2017ris}:
\begin{multline} \label{eq:LALP}
\mathcal{L}_\eta \supset \frac{\partial^\mu \eta}{2} \sum_{\text{f}} \frac{C_{\text{ff}}}{\Lambda}\ \bar{\psi}_\text{f} \gamma_\mu \gamma^5 \psi_{\text{f}} + e^2 \frac{C_{\gamma\gamma}}{\Lambda} \ \eta F_{\mu \nu} \tilde{F}^{\mu \nu} +\\
 \frac{2 e^2}{s_W c_W} \frac{C_{\gamma Z}}{\Lambda}\ \eta F_{\mu \nu} \tilde{Z}^{\mu \nu} + \dots
\end{multline}
where the dots contain couplings to $W^+ W^-$ and $ZZ$, which are irrelevant here. Of the above couplings, the WZW term of Eq.~\eqref{eq:WZW} only generates a non-vanishing
\begin{equation} \label{eq:CgZ}
\frac{C_{\gamma Z}}{\Lambda} = \frac{C_{\rm WZW}}{\Lambda}\,.
\end{equation}
Instead, $C_{\gamma \gamma}$ and $C_{\text{ff}}$ are generated at loop level, and we refer the reader to Ref.~\cite{Bauer:2017ris} for all the necessary details.
As couplings to electrons play a crucial role in the XENON1T signal, we will give some details here about the generation of couplings to fermions:
these loops are divergent, thus the only physical result can be obtained by assuming that the loop is cancelled by a counter-term at a given scale
$\mu$ (that we fix $\mu = f$ in the numerical evaluations). Thus, the loop-induced couplings can be written as
\begin{multline} \label{eq:Cffloop}
\frac{C^{1-loop}_{\text{ff}} (\mu)}{\Lambda} = - \frac{3 \alpha^2 (m_Z)}{2}  \frac{C_{\rm WZW}}{\Lambda} \ln \frac{\mu^2}{m_W^2} \times \\
  \left( \frac{3}{s_W^4} - \frac{4}{c_W^4} (Y_{\text{f}_L}^2 + Y_{\text{f}_R}^2)  \right) \,,
\end{multline}
where f indicates any fermion in the Standard Model, and $Y_{\text{f}_{L/R}}$ are the hypercharges of the left and right-handed fields. 
Numerically, for $f=50$~TeV (and $d_\psi = 4$), we find:
\begin{eqnarray}
\frac{C_{\gamma Z}}{\Lambda} &=&  0.9 \cdot 10^{-7}\, \mbox{GeV}^{-1}\,, \\
\frac{C_{ee}}{\Lambda} &=&  -0.55 \cdot 10^{-8}\, \mbox{GeV}^{-1}\,, \\
\frac{C_{uu}}{\Lambda} &=&  -0.60 \cdot 10^{-8}\, \mbox{GeV}^{-1}\,, \\ 
\frac{C_{dd}}{\Lambda}  &=&  -0.63 \cdot 10^{-8}\, \mbox{GeV}^{-1}\,.
\end{eqnarray}
The scaling with $f$ can be easily inferred from Eqs~\eqref{eq:CWZW} and~\eqref{eq:Cffloop}.
In our benchmark point, the axion coupling $g_{ae}$, relevant for the XENON1T signal, is given by:
\beq
g_{ae} = \frac{m_e C_{ee}}{\Lambda} = - 2.80 \cdot 10^{-12}\,.
\eeq
As the couplings to up and down are approximately equal, the effective coupling to nucleons receives a dominant contribution
from the iso-singlet coupling, thus
\beq
g_{an}^{\rm eff} \approx 0.36\ \frac{m_n}{\Lambda} \frac{C_{uu} + C_{dd}}{2} = - 2.08 \cdot 10^{-9}\,.
\eeq

The coupling to photons, $C_{\gamma \gamma}$, is also generated by loops, and it receives competing contributions both at one-loop from the $W$, and at two-loop from the fermions. Its expression contains several contributions, and we refer the reader to Ref.~\cite{Bauer:2017ris} for more details.
As we are interested in $\eta$ masses below $100$~eV, which is below the $e^+ e^-$ threshold, we find that $C_{\gamma \gamma}$ is suppressed by $m_\eta^2$, thus it is very small. In our benchmark, we find
\beq
g_{a\gamma} = \frac{C_{\gamma \gamma}}{\Lambda} \approx - 10^{-19}\, \mbox{GeV}^{-1}\, \left( \frac{m_\eta}{100~\mbox{eV}} \right)^2\,,
\eeq
which is well below the sensitivity of XENON1T or of any other experiment. This small value also allows to avoid astrophysical and cosmological bounds associated to photon couplings~\cite{Bauer:2017ris}.

\begin{figure}[tb]
\includegraphics[width=0.4\textwidth]{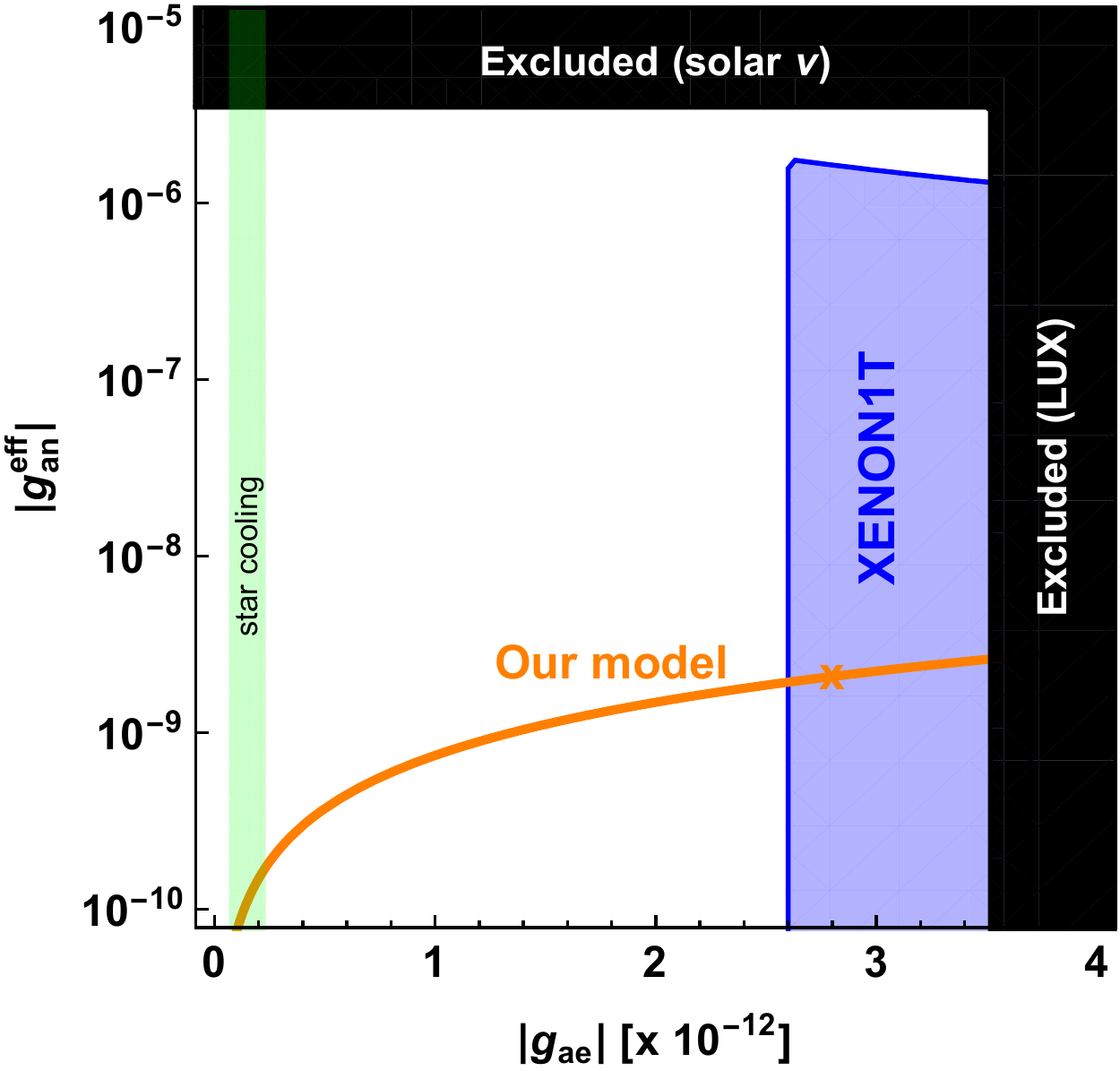}
\caption{In blue, 90\% C.L. region preferred by the XENON1T excess~\cite{Aprile:2020tmw} under the solar axion hypothesis, compared to the prediction of our model in orange. Here we assume $g_{a\gamma} \approx 0$. The black area indicate the excluded are from other direct probes, while the green band could fit the anomalous cooling of stars~\cite{Giannotti:2017hny}. The orange ``x'' indicates the benchmark point $f=50$~TeV.}
\label{fig:Xenon1T}
\end{figure}

The XENON1T collaboration has presented the fit to a generic solar axion model in the parameter space ($|g_{ae}|$, $|g_{ae} g_{an}^{\rm eff}|$, $|g_{ae} g_{a\gamma}|$), providing projections of the allowed pseudo-cuboid on the 3 surfaces. For our model, we can effectively consider $g_{a\gamma} = 0$. This provides the following preferred ranges at 90\% C.L.:
\begin{eqnarray}
& 2.6\cdot 10^{-12} < |g_{ae}| < 3.7 \cdot 10^{-12}\,, & \\
& | g_{ae} g_{an}^{\rm eff}| < 4.5 \cdot 10^{-18}\,, &
\end{eqnarray}
for $m_\eta < 100$~eV. These boundaries define the blue region in Fig.~\ref{fig:Xenon1T}. In the plot, we also report in black the excluded regions by other direct 
experiments: solar neutrino probes, LUX and PandaX-II. The green band represents the $2\sigma$ region that would fit the observed anomalous cooling of stars~\cite{Giannotti:2015kwo,Giannotti:2017hny}, as reported by the XENON1T collaboration.  We will come back to this tension towards the end of this section.

The orange line in the plot shows the predictions for the pseudo-scalar $\eta$ in our model. The shape of the line can be understood in terms of the ratio:
\beq
\left| \frac{g_{an}^{\rm eff}}{g_{ae}} \right| = 743\,,
\eeq
which is independent on $f$. The points inside the 90\% C.L. XENON1T region correspond to
\beq
36 < \frac{f}{\mbox{TeV}} < 55\,.
\eeq
As $f$ is the decay constant of the composite pNGB Higgs, these large values point towards a very fine tuned situation at low energy: a standard measure of the fine tuning is the ratio of $f$ with the electroweak scale~\cite{Panico:2012uw}
\beq
\Delta_{\rm f.t.} = \frac{v_{\rm SM}^2}{f^2} = 2 \div 5 \cdot 10^{-5}\,.
\eeq
In other words, the pNGB Higgs potential at zero temperature needs to be fine-tuned at the level of a part in $10^{5}$. We should remind the reader that this situation, while not optimal, is still a solution to the hierarchy problem between the Planck and electroweak scales: without any new physics,  a tuning at the level of one part in $10^{30}$ would be needed to keep the Higgs scale low. As we will discuss in the next section, fixing $f$ to reproduce the XENON1T excess in our model gives interesting predictions about other phenomena, which could be observed in the future. 
Of course, the first test of this model, as well as of any other explanation, will be in the data taken by the upgraded XENONnT detector~\cite{Aprile:2015uzo} and LZ~\cite{Mount:2017qzi}, which will be able to confirm, or rule as a mere statistical fluctuation, the excess.

Before moving to the predictions, we want to conclude this section discussing two important points that play against the solar axion interpretation of the excess: the potential Tritium contamination and astrophysical data on stellar evolution. The former has been discussed in detail in the XENON1T paper~\cite{Aprile:2020tmw}, and we don't have much to add here. The main conclusion is that adding this background in the fit would reduce the statistical significance of the solar axion explanation from $3.6\sigma$ to $2\sigma$. Yet, only data from the upgrade will be able to confirm or rule out either explanation.
A more crucial issue is presented by stellar data. In fact, it has been already pointed out by the XENON1T collaboration that the preferred values for $g_{ae}$ are at odds with models of stellar evolution, message strengthened in Ref.~\cite{DiLuzio:2020jjp}. The main point is that values of the coupling $g_{ae}$ of the order of $\approx 10^{-12}$ would disrupt the good agreement between stellar evolution models and several available observations, by adding a too fast cooling mechanism via emission of axions. This tension boils down to the fact that $g_{ae}$ should be about one order of magnitude lower than the XENON1T preferred region. However, estimates of astrophysical models are often interpolated from simulations which have many parameters, and the sensitivity to those parameters should be fully taken into account for a fair comparison. Furthermore, observations also have systematic uncertainties to be taken into account. It is thus reasonable that the actual limit could be within one order of magnitude from the quoted one. We, therefore, agree with the pragmatic stand of the XENON1T collaboration in considering the solar axion hypothesis still valid, notwithstanding the tension with stellar data.
Note also that the inclusion of the inverse Primakoff effect at the detector level, as pointed out in Ref.~\cite{Gao:2020wer}, also reduces the tension, however only in models with large coupling to photon, unlike ours.

%%%%%%%%%%%%%%%%%%%%%%%%%%%%%%%%%%%
\section{Predictions}

The beauty of the light pseudo-scalar $\eta$ emerging from the Dark, in composite scenarios, is that its couplings to all Standard Model
particles are ruled by a single parameter: the composite Higgs decay constant $f$. Once it is fixed to reproduce the XENON1T excess, we
can predict several phenomena that may be observable in the near future and confirm, or rule out, this model.

\begin{figure}[tb]
\includegraphics[width=0.4\textwidth]{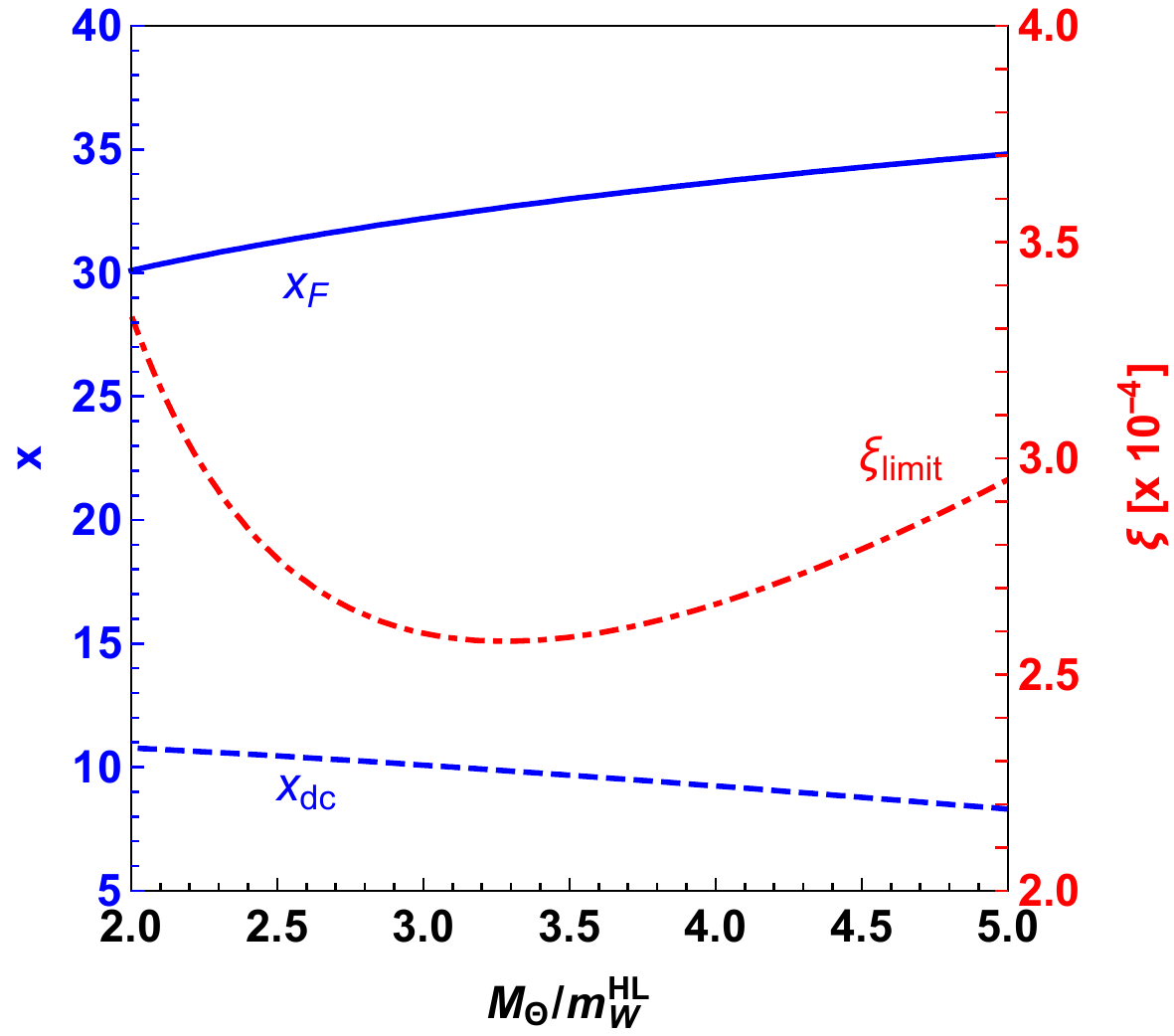}
\caption{In blue, the freeze-out $x_F$ and decoupling $x_{dc}$ parameters for the $\mathbb{Z}_2$--odd states as a function of $M_\Theta/m_W^{HL}$ for $m_W^{\rm HL} = 16$~TeV ($f=50$~TeV). In red, the value of coupling $\xi$ that provides the observed relic density from the asymmetry (smaller values are excluded). }
\label{fig:DMresults}
\end{figure}

\subsubsection*{Dark Matter and gravitational waves}

As the $\eta$ particle emerges from a non-thermal production mechanism for the Dark Matter, we will start with discussing this aspect of the theory.
To do so, we have repeated the analysis presented in~\cite{Cai:2019cow} within the benchmark point $f=50$~TeV. In the Higgsless vacuum, the $W$ mass
is thus given by $m_W^{\rm HL} = m_W f/v_{\rm SM} = 16$~TeV. The dynamics of the Dark Matter evolution depends on two crucial temperatures: a) the
thermal Freeze-out temperature $T_F = M_{\Theta}/x_F$, where $M_{\Theta}$ is the mass of the Dark states in the high-temperature Higgsless vacuum, and b) the decoupling temperature $T_{dc} = M_{\Theta}/x_{dc}$ where the processes transferring the asymmetry between the $\mathbb{Z}_2$--even and $\mathbb{Z}_2$--odd sectors decouple.
The latter is important in fixing the relic abundance of Dark Matter, originating from the asymmetry, while $T_F$ is the largest temperature where the system can start 
drifting away from the Higgsless vacuum alignment, and towards the zero-temperature one (corresponding to the Standard Model). 
The results for $x_F$ and $x_{dc}$ are give by the blue lines in Fig.~\ref{fig:DMresults}, as a function of $M_{\Theta}/m_W^{HL}$.
In red we show the lowest value of the coupling $\xi$ between the Dark Matter and the Higgs boson, where $\xi$ saturating the bound provides the whole relic density via the asymmetry. At zero temperature, this coupling will correspond to the coupling between the Dark Matter and the Standard Model Higgs boson, and values $\xi \approx \mbox{few} \times 10^{-4}$
are required. However, those values are well below the limit from Direct Detection for Dark Matter masses of order of $M_{DM} \approx M_{\Theta} \approx 50$~TeV. 

The Dark Matter production mechanism also requires that the theory undergoes a strong order phase transition, where the strong dynamics confines, at temperatures $T_{\rm HL} \approx f$. This phase transition may generate gravitational waves~\cite{Grojean:2006bp}. The peak frequency of the diffuse gravitational waves mainly depends on the temperature of the phase transition: for instance, from Ref.~\cite{Huang:2020bbe} we can estimate the peak frequency
\begin{equation}
\omega_{\rm peak} \approx 1~\mbox{Hz}\ \frac{1}{v_{\rm w}} \left( \frac{\beta/H_\ast}{1000} \right) \left( \frac{T_{\rm HL}}{100~\mbox{TeV}} \right)\,,
\end{equation}
where $\beta/H_\ast \approx \mathcal{O} (100)$ is a parameter related to the bubble formation, and $v_{\rm w} \approx 1$ is the wall velocity. We thus expect the peak frequency to be in the Hz ballpark, which matches the sensitivity of future detection experiments like Voyager, the Einstein Telescope and the Big Bang Observer (BBO). We leave a more detailed analysis of the gravitational wave spectrum and amplitude for future work.

\subsubsection*{Future collider signals}

The coupling $C_{\gamma Z}$ in Eq.~\eqref{eq:CgZ}  can induce decays of the $Z \to \gamma \eta$, where $\eta$ will appear as missing energy in a detector due to the long lifetime. In fact, for masses in the XENON1T interesting range, $m_\eta < 100$~eV, the dominant decay mode is in neutrinos (as the coupling to photons is negligible), which is suppressed by the neutrino masses and always gives a cosmological stable $\eta$. This exotic decay of the $Z$ could be detectable at a future $e^+ e^-$ collider running at the $Z$ pole, providing a final state with a single mono-energetic photon and missing energy. The branching ratio is given by
\begin{equation}
\Gamma (Z \to \gamma \eta) =\frac{8 \pi \alpha^2 (m_Z)}{3 s_W^2 c_W^2} \left( \frac{C_{\gamma Z}}{\Lambda} \right)^2 m_Z^3\,,
\end{equation}
which gives the following prediction in the range for $f$ preferred by the XENON1T excess:
\begin{equation}
6 < \frac{\mbox{BR} (Z \to \gamma \eta)}{10^{-12}} < 14\,.
\end{equation}
For a future $e^+ e^-$ collider, a Tera-$Z$ run would provide between 6 and 14 events, which may be detectable over the small Standard model background. The analysis would be very simple, and it has already been performed at LEP~\cite{Akers:1994vh,Acciarri:1997im,Abreu:1996vd}, where a bound $\mbox{BR} (Z \to \gamma + \mbox{inv.}) < 10^{-6}$ was established.

\subsubsection*{Flavour changing decays in the down sector, $K$ and $B$ mesons}

The loop-induced couplings of the light pseudo-scalar $\eta$ in Eq.~\eqref{eq:Cffloop} have been computed by extracting the divergent
piece of the loop, which is independent on the mass of the quark in the loop. This is a good approximation for loops containing fermions
much lighter than the $W$ and $Z$ masses, i.e. for all Standard Model fermions except for the top quark. This point has an important
consequence: $W$-loops with the top quark can generate flavour off-diagonal contribution to the coupling of $\eta$ to down-type quarks.
Parametrising them in analogy to Eq.~\eqref{eq:LALP} as
\begin{equation}
\mathcal{L}_\eta \supset \frac{\partial^\mu \eta}{2} \sum_{i\neq j} \frac{C_{d^i d^j}}{\Lambda}\ \bar{\psi}_{d^i} \gamma_\mu \gamma^5 \psi_{d^j}\,, 
\end{equation}
an explicit calculation yields
\begin{equation}
\frac{C_{d^i d^j}}{\Lambda} = - \frac{3 \alpha^2 (m_Z)}{s_W^4} \frac{C_{WZW}}{\Lambda} V^\ast_{td^i} V_{td^j} \frac{1-r_W - \ln r_W}{(1-r_W)^2}\,,
\end{equation}
where $V_{td^i}$ are entries of the CKM matrix and $r_W = m_W^2/m_t^2$.
Interestingly, sizeable off-diagonal couplings only arise in the down quark sector. For the benchmark point $f = 50$~TeV, we find
\begin{eqnarray}
\frac{C_{sd}}{\Lambda} &=& -3.3 \cdot 10^{-13}\, \mbox{GeV}^{-1}\,, \\
\frac{C_{bs}}{\Lambda} &=& -4.8 \cdot 10^{-11}\, \mbox{GeV}^{-1}\,, \\
\frac{C_{bd}}{\Lambda} &=& -1.0 \cdot 10^{-11}\, \mbox{GeV}^{-1}\,,
\end{eqnarray}
where we only report the real part of the coupling.

The first coupling between strange and down quarks is of particular interest, as it can mediate the decay of $K \to \pi + \mbox{inv.}$, where the invisible momentum
is carried away by the $\eta$ pseudo-scalar. This process for neutral Kaons has been recently measured by the KOTO experiment~\cite{Ahn:2018mvc}, which has reported a significant excess over the Standard Model prediction~\cite{Buras:2015qea}:
\begin{eqnarray}
&&\left. \mbox{BR} (K_L \to \pi^0 \rm{inv}) \right|_{\rm KOTO} = 2.1^{+4.1}_{-1.7} \cdot 10^{-9} \,, \\
&&\left. \mbox{BR} (K_L \to \pi^0 \rm{inv}) \right|_{\rm SM} = 3.0 \pm 0.3 \cdot 10^{-11} \,, 
\end{eqnarray}
with errors at 90\% C.L.. To explain this large enhancement with respect to the Standard Model prediction is at odds with the measurement of the
decay of the charged Kaon from NA62 experiment
\begin{eqnarray}
&&\left. \mbox{BR} (K_L \to \pi^0 \rm{inv}) \right|_{\rm NA62} < 2.44 \cdot 10^{-10} \,, \\
&&\left. \mbox{BR} (K_L \to \pi^0 \rm{inv}) \right|_{\rm SM} = 9.11 \pm 0.72 \cdot 10^{-11} \,, 
\end{eqnarray}
as given by the Grossman-Nir bound~\cite{Grossman:1997sk}.

If our model can explain the XENON1T anomaly, then we can also provide a prediction for the New Physics contribution
to these two channels. The calculation can be done following the same procedure as in Ref.~\cite{Fuyuto:2014cya}: we find
\begin{multline}
\Gamma_{K \to \pi + \mbox{inv.}} = \frac{\lambda^{1/2} (1,\hat{m}_\pi^2, \hat{m}_\eta^2) m_K^3}{64 \pi} \left( \frac{C_{sd}}{\Lambda}\right)^2 \times  \\
\left[ (1-\hat{m}_\pi^2) f_+ + \hat{m}_\eta^2 f_- \right]\,,
\end{multline}
where $f_+ \approx 1$ and $f_- \approx -0.28$ are form factors, $\lambda (a,b,c) = a^2 + b^2 + c^2 - 2 a b - 2 a c - 2 b c$ is the usual kinematical function
and $\hat{m}_X = m_X/m_K$. This formula applies both to $K^+$ and to $K_L$, so the difference in branching ratios will only come from the different lifetimes
of the Kaon, thus respecting the Grossman-Nir bound. For the XENON1T preferred region, we find
\begin{eqnarray}
&& \mbox{BR} (K_L \to \pi^0 + \mbox{inv.}) = 0.5 \div 1.2 \cdot 10^{-11}\,, \\
&& \mbox{BR} (K^+ \to \pi^+ + \mbox{inv.}) = 0.12 \div 0.29 \cdot 10^{-11}\,.
\end{eqnarray}
For the $K_L$, the new physics channel enhances the Standard Model prediction by 17\% to 40\%, which is not enough to explain the KOTO anomaly.
Yet, if the precision reaches the level of the Standard Model, and the KOTO anomaly is not confirmed, such enhancement may be a confirmation
of the model. For the charged Kaon, the effect is much smaller, at the percent level.

The couplings involving the bottom quark induce similar decays for the B mesons into pions (via $C_{bd}$) and Kaons (via $C_{bs}$). A calculation similar
to the one done for the Kaons, leads to the predictions
\begin{eqnarray}
&& \mbox{BR} (B \to K + \mbox{inv.}) = 3.4 \div 8.0 \cdot 10^{-9}\,, \\
&& \mbox{BR} (B \to \pi + \mbox{inv.}) = 1.5 \div 3.4 \cdot 10^{-10}\,, 
\end{eqnarray}
where the prediction is roughly the same for both charged and neutral mesons, as the lifetimes are very similar.
The experimental bounds on all those processes are at the level of $10^{-5}$~\cite{Lees:2013kla,Grygier:2017tzo,Tanabashi:2018oca}, while the Standard Model predictions read~\cite{Altmannshofer:2009ma,Buras:2014fpa,Hambrock:2015wka}
\begin{eqnarray}
&& \mbox{BR} (B \to K \nu \nu) = 4.0 \pm 0.5 \cdot 10^{-6}\,, \\
&& \mbox{BR} (B^+ \to \pi^+ \nu \nu) = 2.4 \pm 0.3 \cdot 10^{-7}\,, \\
&& \mbox{BR} (B^0 \to \pi^0 \nu \nu) = 1.2 \pm 0.15 \cdot 10^{-7}\,. 
\end{eqnarray}
The correction deriving from our model, therefore, is expected to be much smaller than the prediction and undetectable in the B meson sector.

%%%%%%%%%%%%%%%%%%%%%%%%%%%%%%%%%%%
\section{Discussion}

The observation of an excess in the electron recoil spectrum for energies below a few keV by the XENON1T has open the tantalising possibility that some
kind of light new physics have been observed. The explanation that best fits the data is that of an axion of solar origin, with masses below $100$~eV. The inclusion 
of an unknown Tritium background, can however reduce the evidence from $3.6\sigma$ to about $2\sigma$, therefore further data are necessary to confirm this
excess as a genuine source of new physics.

The fact that the solar axion hypothesis seems at odds with solar cooling models~\cite{DiLuzio:2020jjp}, has ignited the community in finding alternative explanations: 
for instance, the signal could be due to a photophobic warm-Dark-Matter axion~\cite{Takahashi:2020bpq} or another boson produced in the sun~\cite{Chen:2020gcl,An:2020bxd,Bloch:2020uzh,Budnik:2020nwz}.  The possibility of exotic solar neutrino interactions, induced by a light mediator, has also been considered by several authors~\cite{Boehm:2020ltd,Bally:2020yid,AristizabalSierra:2020edu,Khan:2020vaf,Jho:2020sku}, as well as a variety of mechanism that could boost a light Dark Matter candidate~\cite{Fornal:2020npv,Harigaya:2020ckz,Cao:2020bwd,Primulando:2020rdk,Lee:2020wmh,Jho:2020sku,Baryakhtar:2020rwy,Bramante:2020zos,Bloch:2020uzh}, with some consensus around models with two states split by only a few keV. 
Finally, the possibility of a non-thermal Hidden photon Dark Matter~\cite{Alonso-Alvarez:2020cdv,Choi:2020udy,Nakayama:2020ikz}, of photons emitted by Dark Matter~\cite{Bell:2020bes,Paz:2020pbc,Dey:2020sai}, mirror symmetry~\cite{Zu:2020idx} or U(1)-2HDM~\cite{Lindner:2020kko} have been considered.

In this letter, we stand for the solar axion explanation and find that a good candidate was predicted by a new mechanism of Dark Matter production we proposed in Ref.~\cite{Cai:2019cow}. 
We then propose a composite Higgs scenario as the most constrained scenario. 
Besides the mass, which is expected to be very light and can be in the range relevant for the XENON1T excess, $m_\eta < 100$~eV, the 
couplings of the ALP to Standard Model particles are all ruled by a single parameter: the Higgs decay constant $f$. 
Once $f$ is fixed in the range $36-55$~TeV to explain the XENON1T excess, we can make several predictions which are testable at future experiments.

First, the large value of the Higgs decay constant implies a severe fine tuning in the Higgs potential, of the order of a few parts in $10^5$: this still being a solution
to the hierarchy problem, this implies that only a future 100 TeV hadron collider may be able to directly probe the composite nature of the Higgs boson. Furthermore, the Dark Matter candidate is predicted to have masses in the $50$~TeV range and small couplings to the Higgs boson, which makes its direct detection very challenging.
On the bright side, the necessity for a strong first order phase transition at a temperature close to $f$ implies the production of gravitational waves with a typical peak
frequency around the Hz, thus being potentially detectable by future experiments like Voyager, the Einstein Telescope and the Big Bang Observer (BBO).

The lightness of the pseudo-scalar also implies the presence of exotic decays of Standard Model particles: we identified two interesting channels: $Z \to \gamma + \mbox{inv.}$ and $K_L \to \pi^0 + \mbox{inv.}$. The former is predicted with rates between $6 \div 14 \cdot 10^{-12}$, which means that a handful of events are predicted at a future $e^+ e^-$ collider running in the Tera- $Z$ mode. Due to the very low Standard Model background, and the mono-chromatic nature of the photon, this process may be observable, and we leave a detailed analysis for future work.
The latter process is due to flavour changing couplings of the ALP to down-type quarks, due to the presence of the heavy top in the loop. The fact that such effects are only expected for down-type quarks is a nice consequence of the loop nature of these couplings. We predict that the rate $K_L \to \pi^0 + \mbox{inv.}$ receives a $+17 \div 40$\% enhancement due to the new physics channel. This is not enough to explain the recent KOTO anomaly, but it could be detected once experimental precision reaches the level of the Standard Model prediction. Other channels, like $K^+ \to \pi^+ + \mbox{inv.}$, $B_0 \to K_0/\pi^0 +  \mbox{inv.}$ and $B^+ \to K^+/\pi^+ +  \mbox{inv.}$, all receive much smaller contributions.

In conclusion, we have identified a very predictive scenario of composite Higgs and Dark Matter where the XENON1T excess can be explained by a very light ALP, at the price of a severe fine tuning in the Higgs potential. Due to the limited number of free parameters in the ALP sector, we made several testable prediction, from gravitational waves of frequencies in the Hz ballpark, to new mono-photon decays of the $Z$ which will give a few events at a future Tera-$Z$ collider, to an enhancement in the Kaon decay $K_L \to \pi^0 + \mbox{inv}$. Finally, due to the compositeness scale being in the $36 \div 55$~TeV range, only a $100$~TeV collider may be able to directly produce new heavy resonances connected to the composite nature of the Higgs boson and of the Dark Matter. \\

\section*{Acknowledgements}
GC acknowledges partial support from the Labex-LIO (Lyon Institute of Origins) under grant ANR-10-LABX-66 (Agence Nationale pour la Recherche), and FRAMA (FR3127, F\'ed\'eration de Recherche ``Andr\'e Marie Amp\`ere''). 
CC and HHZ are supported by the National Natural Science Foundation of China (NSFC) under Grant Nos. 11875327 and 11905300, the China Postdoctoral Science Foundation under Grant No. 2018M643282, the Natural Science Foundation of Guangdong Province under Grant No. 2016A030313313, the Fundamental Research Funds for the Central Universities, and the Sun Yat-Sen University Science Foundation.
MTF and MR acknowledge partial funding from The Council For Independent Research, grant number DFF 6108-00623. The CP3-Origins centre is partially funded by the Danish National Research Foundation, grant number DNRF90.
GC, CC and HHZ also acknowledge support from the China-France LIA FCPPL.
GC also thanks the Sun Yat-Sen University for hospitality during the completion of this project.

%\appendix
%
%\section{Appendix}
%

\bibliographystyle{JHEP-2-2}

\bibliography{bibHiggsfDark.bib}

\end{document}